\documentclass[11pt,a4paper]{article}

\usepackage[utf8]{inputenc}
\usepackage[T1]{fontenc}
\usepackage[margin=2.5cm]{geometry}
\usepackage{amsmath,amssymb,amsfonts,amsthm}
\usepackage{mathpazo}
\usepackage{graphicx}
\usepackage{xcolor}
\usepackage{booktabs}
\usepackage{tabularx}
\usepackage{float}
\usepackage{enumitem}
\usepackage[colorlinks=true,linkcolor=blue,citecolor=blue,urlcolor=blue]{hyperref}

\newtheorem{Theorem}{Theorem}
\newtheorem{Proposition}{Proposition}
\newtheorem{Corollary}{Corollary}
\theoremstyle{definition}
\newtheorem{Definition}{Definition}
\theoremstyle{remark}
\newtheorem{Remark}{Remark}

\title{On the Geometric Structure of Hyperbolic Clifford Bundles\\ and Associated Spin Groups}

\author{Eduardo Notte-Cuello\\[4pt]
\small Departamento de Matem\'{a}ticas, Universidad de La Serena (ULS),\\
\small La Serena 1720256, Chile\\
\small \texttt{enotte@userena.cl}}

\date{}

\begin{document}

\maketitle

\begin{abstract}
In this paper, we present the spinor structure associated with the Hyperbolic
Clifford algebra of a real $n$-dimensional vector space $V$, which is denoted
by $\mathcal{C}l\left(H_{V}\right)$. Unlike the standard Clifford algebra, the
Hyperbolic Clifford algebra $\mathcal{C}l(H_V)$ simultaneously accommodates both
multiforms and multivectors in a single algebraic structure, making it the
natural framework---known as the ``mother algebra''---for the study of
superfields in theoretical physics and for generalizing the Clifford bundle
formalism to hyperbolic structures arising in gravitational theories. The
orthogonal groups and orthogonal transformations associated to the hyperbolic
space $H_{V}$ are presented. The Clifford--Lipschitz group and the Pin and Spin
groups associated with $\mathcal{C}l\left(H_{V}\right)$ are defined. Then, the
frame bundle and spinor structure associated to Hyperbolic Clifford algebra is
derived.
\end{abstract}



\section{Introduction}

It is known that in General Relativity, the gravitational field is understood
as an element or facet of a geometrical structure of the spacetime manifold;
actually, we have that each gravitational field generated by a given matter
distribution is modeled by a pentuple $\left(M,g,D,\tau_{g},\uparrow\right)$,
where $M$ is a $4$-dimensional manifold, $g\in\sec T_{2}^{0}M$ is
a Lorentzian metric on $M$, $D$ is the Levi--Civita connection of $g$,
$\tau_{g}\in\sec{\textstyle\bigwedge\nolimits^{4}}T^{\ast}M$ is the volume
element that defines the spatial orientability of $M$ and $\uparrow$ means
that $M$ is time oriented. Here the particles are described by triples
$\left\langle \left(m,q\right),S,\sigma\right\rangle$ where $m$ is the
particle's mass, $q$ the electric charge, $\sigma$ is the world line and $S$
is the spin of the particle, which is characterized in the Clifford bundle
formalism; for details see, e.g.,~\cite{Rodrigues1,Sachs}.

Clifford bundle formalism is a theory that has proven to be a convincing
alternative for the study of the gravitational field. For example, in
\cite{Rodrigues3} Einstein's gravitational theory is established with this
formalism, which suggests that the gravitational field is in
Minkowski spacetime. In~\cite{Notte2}, by using the Clifford bundle formalism,
a Lagrangian theory of the Yang--Mills type for the gravitational field in
Minkowski spacetime is developed; there, it is shown how two simple
hypotheses permit the interpretation of the formalism in terms of effective
Lorentzian or teleparallel geometries, and in the case of a Lorentzian geometry
interpretation of the theory, the field equations are shown to be equivalent to
Einstein's equations. In~\cite{fr2010}, using this same formalism, a theory of
the gravitational field is presented. This field is represented by a
$(1,1)$-extensor field $h$ that describes a plastic distortion of the Lorentz
vacuum due to the presence of matter. This theory allows the introduction of
different types of parallelism rules in the world manifold, which can be
interpreted as distortions of the parallelism structure of Minkowski spacetime.

For the study of the theory of gravitation and other aspects of theoretical
physics, the Clifford bundle formalism can be generalized over a Clifford
algebra associated with a hyperbolic space, where the elements of the algebra
now consist of both multiforms and multivectors. The simultaneous treatment of
multiforms and multivectors in a single algebra is physically motivated by the
need to provide a unified description of fields and their dual counterparts, as
required for the study of superfields and for formulations of gravitational
theories that go beyond the standard Clifford bundle approach. This general
algebraic structure is called the Hyperbolic Clifford algebra and the aim of
this work is to introduce a spinor structure into this theory, which is
fundamental for the study of various topics in Theoretical Physics.

The Hyperbolic Clifford algebra of a real $n$-dimensional vector space $V$,
oriented to the study of superfields in theoretical physics, was first
introduced in~\cite{Rodrigues2}, where an incipient development of the
hyperbolic structure describing these superfields was presented, and later in
\cite{qro7} the same authors presented a detailed development of the hyperbolic
space and the Hyperbolic Clifford algebra, oriented to the study of
superfields. In this last work, the authors recall the construction of a
hyperbolic space $H_{V}=\left(V\oplus V^{\ast},\left\langle ,\right\rangle\right)$
endowed with a non-degenerate bilinear form $\left\langle ,\right\rangle$ and
they show, among other things, that for a symmetric bilinear form $b$ of
arbitrary signature, there exists a relation between $H_{V}$ and the exterior
direct sums of the hyperbolic spaces associated with the pairs $(V,b)$ and
$(V,-b)$; they define the Clifford algebra $\mathcal{C}l\left(H_{V}\right)$ of
\emph{multivecfors} associated with the hyperbolic space $H_{V}$ and present
several formulas that are similar to those of a Clifford algebra of
multivectors on $V$, which are fundamental for the development of the theory.
Later in~\cite{Notte1}, we presented an extensive review of
$\mathcal{C}l\left(H_{V}\right)$ and a review of the differential
structure of the Hyperbolic Clifford algebra, where we exposed the properties
of the \emph{duality product} of multivectors and multiforms and the theory of
$k$ multivector and $l$ multiform variables multivector extensors over $V$.
Also in that work, we studied the theory of parallelism structures on an
arbitrary smooth manifold $M$, and the concepts of covariant derivatives,
deformed derivatives and relative covariant derivatives of multivector,
multiform and extensor fields. All these theories and concepts are
fundamental to the study of various physical theories; see, e.g.,~\cite{fr2010}.
However, none of the previous works have carried out a formal development of
the spinor structure of the Hyperbolic Clifford algebra, which is necessary to
characterize the properties of the particles, as well as the topics developed
in the context of the Clifford algebra of multivectors or multiforms. The
present paper fills this gap.

Unlike standard Clifford algebras, which deal exclusively with multivectors,
the hyperbolic Clifford algebra $\mathcal{C}\ell(H_V)$ provides a unified
framework that simultaneously accommodates both multivectors (elements of
$\bigwedge V$) and multiforms (elements of $\bigwedge V^*$) in a single
algebraic structure. This unified treatment is essential for the study of
superfields in theoretical physics and for developing gravitational theories
that require symmetric treatment of fields and their dual counterparts. It can
be shown (see, e.g.,~\cite{qro7}) that $\bigwedge V^*$ is a minimal left ideal
of the hyperbolic Clifford algebra, i.e., $\bigwedge V^* = \mathcal{C}\ell(H(M))\theta^*$,
and the elements of $\mathcal{C}\ell(H(M))\theta^*$ are Witten superfields.

The primary contributions of this paper are as follows:

\begin{enumerate}
\item[(a)] Formal development of spinor structures for hyperbolic Clifford
algebras, extending results from standard Clifford algebra theory to this
generalized hyperbolic setting.

\item[(b)] Definition and analysis of hyperbolic orthogonal groups,
Clifford--Lipschitz groups, Pin and Spin groups associated with
$\mathcal{C}\ell(H_V)$, with explicit surjective homomorphisms to the
respective orthogonal transformation groups.

\item[(c)] Construction of hyperbolic Clifford bundles on Lorentzian manifolds,
establishing the associated vector bundle structure and reduction in the
structure group.

\item[(d)] Unified formalism for studying gravitational fields and particles
in theories that require simultaneous treatment of fields and their dual
counterparts.
\end{enumerate}

In this paper, we present the construction of the frame bundle associated with
the Hyperbolic Clifford algebra and the Hyperbolic Clifford bundle of a real
$n$-dimensional vector space $V$. We show that this is a vector bundle
associated with the principal bundle of orthonormal frames. In
Section~\ref{sec2}, we present a brief review of the Hyperbolic space $H_{V}$
and the Hyperbolic Clifford algebra $\mathcal{C}l\left(H_{V}\right)$ associated
with this space, as well as some identities necessary for the development of
the theory (see also Table~\ref{tab:comparison}). In Section~\ref{sec3} we study
the groups associated with $\mathcal{C}l\left(H_{V}\right)$, namely the
orthogonal groups and symmetries in the $H_{V}$ space as well as the
Clifford--Lipschitz groups and the Pin and Spin groups associated with
$\mathcal{C}l\left(H_{V}\right)$. Finally, in Section~\ref{sec4}, we present
the construction of the frame bundle associated with
$\mathcal{C}l\left(H_{V}\right)$ and the Hyperbolic Clifford bundle, and then we
derive the spinor structure for $\mathcal{C}l\left(H_{V}\right)$.

\section{Hyperbolic Spaces}\label{sec2}

\subsection{Definition and Basic Properties}

In this section, we recall some definitions and basic properties of
Hyperbolic space. Let $V,V^{\ast}$ be a pair of dual $n$-dimensional vector
spaces over the real field $\mathbb{R}$ and $V\oplus V^{\ast}$ the exterior
direct sum of the vector spaces $V$ and $V^{\ast}$; for more details see
\cite{Notte1,qro7}. We recall that a hyperbolic structure over $V$ is the pair
\[
H_{V}=\left(V\oplus V^{\ast},\left\langle ,\right\rangle \right)
\]
where $\left\langle ,\right\rangle$ is the non-degenerate symmetric bilinear
form of index $n$ defined by\footnote{This non-degenerate symmetric bilinear
form was first introduced in~\cite{Oziewicz}.}
\begin{equation}
\left\langle \mathbf{x},\mathbf{y}\right\rangle =x^{\ast}\left(y_{\ast}\right)
+y^{\ast}\left(x_{\ast}\right) \label{a1}
\end{equation}
for all $\mathbf{x}=x_{\ast}\oplus x^{\ast},\mathbf{y}=y_{\ast}\oplus y^{\ast}\in H_{V}$.

\begin{Remark}
In the Witt basis $\{e_1,\ldots,e_n,\theta^1,\ldots,\theta^n\}$ of $H_V$, the
matrix representation of the bilinear form $\langle\cdot,\cdot\rangle$ is
\[
\begin{pmatrix} 0 & I_n \\ I_n & 0 \end{pmatrix},
\]
where $I_n$ is the $n\times n$ identity matrix. This makes it explicit that the
form is neutral (i.e., of signature $(n,n)$), and that the subspaces $V$ and
$V^*$ are totally isotropic, since the diagonal blocks are zero.
\end{Remark}

The elements of $H_{V}$ are referred to as \textit{vecfors}. A vecfor
$\mathbf{x}=x_{\ast}\oplus x^{\ast}\in H_{V}$ is positive if
$x^{\ast}\left(x_{\ast}\right)>0$, null if $x^{\ast}\left(x_{\ast}\right)=0$ and
negative if $x^{\ast}\left(x_{\ast}\right)<0$. If $x^{\ast}\left(x_{\ast}\right)=1$
we say that $\mathbf{x}=x_{\ast}\oplus x^{\ast}$ is a unit vecfor.

Note that the spaces $V$ and $V^{\ast}$ are maximal totally isotropic subspaces
of $H_{V}$. In reality these spaces can be identified in $H_{V}$ under the
image of the inclusions $i_{\ast}:V\rightarrow V\oplus V^{\ast}$,
$x_{\ast}\mapsto i_{\ast}x_{\ast}=x_{\ast}\oplus0\equiv x_{\ast}$ and
$i^{\ast}:V^{\ast}\rightarrow V\oplus V^{\ast}$,
$x^{\ast}\mapsto i^{\ast}x^{\ast}=0\oplus x^{\ast}\equiv x^{\ast}$; then from
(\ref{a1}) we have
\[
\left\langle i_{\ast}x_{\ast},i_{\ast}y_{\ast}\right\rangle =\left\langle
x_{\ast},y_{\ast}\right\rangle =0,\quad\left\langle i^{\ast}x^{\ast},i^{\ast}
y^{\ast}\right\rangle =\left\langle x^{\ast},y^{\ast}\right\rangle
=0,\quad\left\langle i^{\ast}x^{\ast},i_{\ast}y_{\ast}\right\rangle =x^{\ast}\left(y_{\ast}\right)
\]
for all $x_{\ast},y_{\ast}\in V\subset V\oplus V^{\ast}$ and
$x^{\ast},y^{\ast}\in V^{\ast}\subset V\oplus V^{\ast}$.

More generally, to each subspace $S\subset V$ (or analogously,
$S^{\prime}\subset V^{\ast}$) we can associate a maximal totally isotropic
subspace as follows: define the null subspace $S^{\prime}\subset V^{\ast}$ of
$S\subset V$ by
\[
S^{\prime}=\left\{ x^{\ast}\in V^{\ast}:x^{\ast}\left(y_{\ast}\right)
=0,\ \forall y_{\ast}\in V\right\}.
\]
Note that $S^{\prime}$ is a subspace of $V^{\ast}$. Furthermore, it is easy to
prove that if $\dim V=n$, then $\dim\left(S^{\prime}\right)=n-\dim S$;
then the $n$-dimensional vector subspace $S\oplus S^{\prime}$ of $H_{V}$ is a
maximal totally isotropic subspace of $H_{V}$.

On the other hand, if we consider the basis $\left\{ e_{1},\cdots,e_{n}\right\}$
of $V$ and $\left\{ \theta^{1},\cdots,\theta^{n}\right\}$ of $V^{\ast}$, then in
$H_{V}$ we have that
\[
\left\langle e_{i},e_{j}\right\rangle =0,\quad\left\langle \theta^{i},
\theta^{j}\right\rangle =0\quad\text{and}\quad\left\langle \theta^{i},
e_{j}\right\rangle =\delta_{j}^{i}.
\]
To obtain an orthogonal basis $\left\{ \mathbf{\sigma}_{1},\cdots,\mathbf{\sigma}_{2n}\right\}$
of $H_{V}$, we consider a Witt basis
$\left\{ e_{1},\ldots,e_{n},\theta^{1},\ldots,\theta^{n}\right\}$ and we
define the elements of this basis as follows:
\begin{equation}
\mathbf{\sigma}_{k}=\frac{1}{\sqrt{2}}\left(e_{k}\oplus\theta^{k}\right)
\quad\text{and}\quad\mathbf{\sigma}_{n+k}=\frac{1}{\sqrt{2}}\left(
\overline{e}_{k}\oplus\theta^{k}\right) \label{a2}
\end{equation}
where $\overline{e}_{k}=-e_{k}$, $k=1,\ldots,n$. Thus, for $k,l=1,\ldots,n$ we have
\[
\left\langle \mathbf{\sigma}_{k},\mathbf{\sigma}_{l}\right\rangle =\delta
_{kl},\quad\left\langle \mathbf{\sigma}_{n+k},\mathbf{\sigma}_{n+l}
\right\rangle =-\delta_{kl}\quad\text{and}\quad\left\langle \mathbf{\sigma}
_{k},\mathbf{\sigma}_{n+l}\right\rangle =0.
\]
It is important to note here that the basis vectors $\mathbf{\sigma}_{n+l}$ are
obtained from the $\mathbf{\sigma}_{k}$ through the involution
$\mathbf{x}=x_{\ast}\oplus x^{\ast}\mapsto\overline{\mathbf{x}}=(-x_{\ast})\oplus x^{\ast}$,
where $\overline{\mathbf{x}}$ is the (hyperbolic) conjugate of the vecfor $\mathbf{x}$.

A very important result concerning the theory of hyperbolic spaces is the following.

\begin{Proposition}
Given an arbitrary non-degenerate symmetric bilinear form $b$ on $V$, there is
an isomorphism
\[
H_{V}\approx\left(V,b\right)\oplus\left(V,-b\right)=H_{bV}.
\]
\end{Proposition}

This result can be found in~\cite{Knus,qro7}. The explicit
isomorphism involved in this proposition is given by
\begin{equation}
\chi_{b}:H_{V}\longrightarrow H_{bV},\quad\chi_{b}\left(x_{\ast}+x^{\ast}
\right)=x_{+}+x_{-} \label{a3}
\end{equation}
where
\[
x_{\pm}=\frac{1}{\sqrt{2}}\left(b^{\ast}x^{\ast}\pm x_{\ast}\right)
\]
with $b^{\ast}x^{\ast}\equiv b^{\ast}\left(x^{\ast},\cdot\right)$ for all
$x^{\ast}\in V^{\ast}$, where $b^{\ast}$ is a bilinear form on $V^{\ast}$,
reciprocal of $b$, i.e., $b^{ik}b_{kj}=\delta_{j}^{i}$, $b_{ij}=b\left(e_{i},e_{j}\right)$,
$b^{ik}=b^{\ast}\left(\theta^{i},\theta^{k}\right)$.
For details on how this isomorphism operates on a basis of $H_{V}$, see
\cite{qro7}. In particular, we can take the vector space $V=\mathbb{R}^{p,q}$
with the bilinear form $b$ of signature $\left(p,q\right)$.

\subsection{Exterior Algebra and Contractions of a Hyperbolic Space}

The Grassmann algebra $\bigwedge H_{V}$ of the hyperbolic structure $H_{V}$ is
the pair
\[
\bigwedge H_{V}=\left(\bigwedge\left(V\oplus V^{\ast}\right),\left\langle ,\right\rangle \right)
\]
where
\[
\bigwedge\left(V\oplus V^{\ast}\right)=\sum\limits_{r=0}^{2n}
\bigwedge\nolimits^{r}\left(V\oplus V^{\ast}\right)
\]
is the exterior algebra of $V\oplus V^{\ast}$ and $\left\langle ,\right\rangle$
is the canonical bilinear form on $\bigwedge\left(V\oplus V^{\ast}\right)$
induced by the bilinear form $\left\langle ,\right\rangle$ of $H_{V}$, and
it is extended by linearity and orthogonality to all of the algebra
$\bigwedge H_{V}$. Of course, due to the isomorphisms
$H_{V}^{\ast}\simeq H_{V^{\ast}}\simeq H_{V}$, we have
\[
\bigwedge H_{V}^{\ast}\simeq\bigwedge H_{V^{\ast}}\simeq\bigwedge H_{V},
\]
and it follows that
\[
\bigwedge H_{V}\simeq\left(\bigwedge H_{V}\right)^{\ast},
\]
i.e., $\bigwedge H_{V}$ is itself an auto-dual space. The elements of
$\bigwedge H_{V}$ are called \textit{multivecfors}.

Grade involution, reversion and conjugation in the algebra $\bigwedge H_{V}$
are defined as usual; see, e.g.,~\cite{qro7,Rodrigues1}. For homogeneous
multivecfors $\mathbf{u}\in\bigwedge^{r}H_{V}$,
\[
\widehat{\mathbf{u}}=\left(-1\right)^{r}\mathbf{u},\quad\widetilde
{\mathbf{u}}=\left(-1\right)^{\frac{1}{2}r\left(r-1\right)}\mathbf{u},
\quad\overline{\mathbf{u}}=\widehat{\widetilde{\mathbf{u}}}=\left(-1\right)
^{\frac{1}{2}r\left(r+1\right)}\mathbf{u}
\]
and we recall that every element $\mathbf{u}\in\bigwedge H_{V}$ is uniquely
decomposed into a sum of its even and odd parts, i.e.,
$\mathbf{u}=\mathbf{u}_{+}+\mathbf{u}_{-}$, where
\[
\mathbf{u}_{+}=\frac{1}{2}\left(\mathbf{u}+\widehat{\mathbf{u}}\right)
\quad\text{and}\quad\mathbf{u}_{-}=\frac{1}{2}\left(\mathbf{u}-\widehat{\mathbf{u}}\right).
\]

\begin{Remark}
The grade involution of $\mathbf{x}\in H_{V}$, denoted by $\widehat{\mathbf{x}}$,
is also denoted by the map $\alpha:H_{V}\rightarrow H_{V}$, with
$\alpha\left(\mathbf{x}\right)=\widehat{\mathbf{x}}=\left(-x_{\ast}\right)\oplus x^{\ast}$.
In general, every Clifford algebra admits a unique canonical automorphism or
involutionary automorphism
$\alpha:\mathcal{C\ell}\left(\mathbb{V}\right)\rightarrow\mathcal{C\ell}\left(\mathbb{V}\right)$
that maps every odd element $x\in\mathcal{C\ell}^{1}\left(\mathbb{V}\right)$ to
$-x$, and which satisfies the following properties:
\[
\alpha\circ\alpha=id;\qquad\alpha\left(xy\right)=\alpha\left(x\right)
\alpha\left(y\right);\qquad\alpha\left(v\right)=-v,\ \forall v\in\mathbb{V}.
\]
\end{Remark}

The spaces $\bigwedge V$ and $\bigwedge V^{\ast}$ are identified with their
images in $\bigwedge H_{V}$ under the homomorphisms
$i^{w}:\bigwedge V\rightarrow\bigwedge H_{V}$ defined by
\[
i^{w}\left(x_{1\ast}\wedge\ldots\wedge x_{r\ast}\right)=\left(x_{1\ast}
\oplus0\right)\wedge\ldots\wedge\left(x_{r\ast}\oplus0\right)\equiv
x_{1\ast}\wedge\ldots\wedge x_{r\ast}
\]
and $i_{w}:\bigwedge V^{\ast}\rightarrow\bigwedge H_{V}$ defined by
\[
i_{w}\left(x_{1}^{\ast}\wedge\ldots\wedge x_{r}^{\ast}\right)=\left(0\oplus
x_{1}^{\ast}\right)\wedge\ldots\wedge\left(0\oplus x_{r}^{\ast}\right)
\equiv x_{1}^{\ast}\wedge\ldots\wedge x_{r}^{\ast}.
\]
Then, for $u_{\ast},v_{\ast}\in\bigwedge V\subset\bigwedge H_{V}$ and
$u^{\ast},v^{\ast}\in\bigwedge V^{\ast}\subset\bigwedge H_{V}$, we have
\begin{equation}
\left\langle u_{\ast},v_{\ast}\right\rangle =0,\quad\left\langle u^{\ast},
v^{\ast}\right\rangle =0\quad\text{and}\quad\left\langle u^{\ast},v_{\ast}
\right\rangle =u^{\ast}\left(v_{\ast}\right). \label{a5}
\end{equation}
Thus, $\bigwedge V$ and $\bigwedge V^{\ast}$ are totally isotropic subspaces of
$\bigwedge H_{V}$. But they are no longer maximal: $\left\langle ,\right\rangle$
being neutral, the dimension of a maximal totally isotropic subspace is
$2^{2n-1}$, whereas $\dim\bigwedge V=\dim\bigwedge V^{\ast}=2^{n}$. Indeed, the
bilinear form $\langle\cdot,\cdot\rangle$ on $\bigwedge H_V$ is neutral, i.e.,
it has signature $(2^{2n-1}, 2^{2n-1})$. For a neutral form on a
$2^{2n}$-dimensional space, the dimension of a maximal totally isotropic
subspace equals exactly half the total dimension, i.e., $2^{2n-1}$. Since
$\dim\bigwedge V = 2^n < 2^{2n-1}$ for $n\geq 2$, the subspaces $\bigwedge V$
and $\bigwedge V^*$ are totally isotropic but not maximal. For elements
$\mathbf{u}=u_{\ast}\wedge u^{\ast},\mathbf{v}=v_{\ast}\wedge v^{\ast}\in\bigwedge H_{V}$
with $u_{\ast}\in\bigwedge^{r}V$, $u^{\ast}\in\bigwedge^{s}V^{\ast}$,
$v_{\ast}\in\bigwedge^{s}V$ and $v^{\ast}\in\bigwedge^{r}V^{\ast}$ it holds
\[
\left\langle \mathbf{u},\mathbf{v}\right\rangle =\left(-1\right)
^{rs}u^{\ast}\left(v_{\ast}\right)v^{\ast}\left(u_{\ast}\right).
\]

It is also very important to recall that the volume element
$\mathbf{\sigma}=\mathbf{\sigma}_{1}\wedge\ldots\wedge\mathbf{\sigma}_{2n}\in\bigwedge\nolimits^{2n}H_{V}$
with $\mathbf{\sigma}_{x}\neq0$ $\forall x\in M$ provides an orientation to the
hyperbolic space $H_{V}$; here
$\left\{ \mathbf{\sigma}_{1},\cdots,\mathbf{\sigma}_{2n}\right\}$ is the
orthonormal basis of $H_{V}$ naturally associated with the dual bases
$\left\{ e_{1},\cdots,e_{n}\right\}$ of $V$ and
$\left\{ \theta^{1},\cdots,\theta^{n}\right\}$ of $V^{\ast}$.

On the other hand, a left contraction
$\lrcorner:{\textstyle\bigwedge}H_{V}\times{\textstyle\bigwedge}H_{V}\rightarrow{\textstyle\bigwedge}H_{V}$
and a right contraction
$\llcorner:{\textstyle\bigwedge}H_{V}\times{\textstyle\bigwedge}H_{V}\rightarrow{\textstyle\bigwedge}H_{V}$
are introduced in the algebra ${\textstyle\bigwedge}H_{V}$ in the usual way
(see, e.g.,~\cite{qro7,Rodrigues1}), i.e., by
\[
\left\langle \mathbf{u}\lrcorner\mathbf{v},\mathbf{w}\right\rangle
=\left\langle \mathbf{v},\widetilde{\mathbf{u}}\wedge\mathbf{w}\right\rangle
\quad\text{and}\quad\left\langle \mathbf{v}\llcorner\mathbf{u},\mathbf{w}
\right\rangle =\left\langle \mathbf{v},\mathbf{w}\wedge\widetilde{\mathbf{u}}\right\rangle,
\]
for all $\mathbf{u},\mathbf{v},\mathbf{w}\in{\textstyle\bigwedge}H_{V}$. The
general properties, for all $\mathbf{x},\mathbf{y}\in H_{V}$, for all
$\mathbf{u},\mathbf{v},\mathbf{w}\in{\textstyle\bigwedge}H_{V}$ and
$\mathbf{\sigma}\in\bigwedge\nolimits^{2n}H_{V}$, can be seen in
\cite{qro7,Notte1}. Moreover, from Equation~(\ref{a5}) we obtain
\begin{equation}
u_{\ast}\lrcorner v_{\ast}=u_{\ast}\llcorner v_{\ast}=0,\quad u^{\ast}
\lrcorner v^{\ast}=u^{\ast}\llcorner v^{\ast}=0 \label{a6}
\end{equation}
for all $u_{\ast},v_{\ast}\in{\textstyle\bigwedge}V$ and
$u^{\ast},v^{\ast}\in{\textstyle\bigwedge}V^{\ast}$, so that for elements of the
form $\mathbf{u}=u_{\ast}\wedge u^{\ast}$ and $\mathbf{x}=x_{\ast}\oplus x^{\ast}$,
it holds
\[
\begin{array}[c]{c}
\mathbf{x}\lrcorner\mathbf{u}=\left(x^{\ast}\lrcorner u_{\ast}\right)
\wedge u^{\ast}+\widehat{u}_{\ast}\wedge\left(x_{\ast}\lrcorner u^{\ast}\right)\medskip\\
\mathbf{u}\llcorner\mathbf{x}=u_{\ast}\wedge\left(u^{\ast}\llcorner x_{\ast}
\right)+\left(x^{\ast}\lrcorner u_{\ast}\right)\wedge\widehat{u}^{\ast}
\end{array}
\]
For more details about the properties of the left and right contractions, see,
e.g.,~\cite{Lounesto,Rodrigues1}.

\subsection{Clifford Algebra of a Hyperbolic Space}

Introduce in ${\displaystyle\bigwedge}H_{V}$ the Clifford product of a vecfor
$\mathbf{x}\in H_{V}$ by an element $\mathbf{u}\in{\textstyle\bigwedge}H_{V}$ by
\[
\mathbf{xu}=\mathbf{x}\lrcorner\mathbf{u}+\mathbf{x}\wedge\mathbf{u}
\]
and extend this product by linearity and associativity to all of the space
${\textstyle\bigwedge}H_{V}$; the general properties of the Clifford product
can be seen, e.g., in~\cite{Rodrigues1}. The resulting algebra is isomorphic to
the Clifford algebra $\mathcal{C\ell}(H_{V})$ of the hyperbolic structure
$H_{V}$ and will thereby be identified with it and is called the \emph{mother}
algebra (or the hyperbolic Clifford algebra) of the vector space $V$. The even
and odd subspaces of $\mathcal{C\ell}(H_{V})$ will be denoted, respectively, by
$\mathcal{C\ell}^{(0)}(H_{V})$ and $\mathcal{C\ell}^{(1)}(H_{V})$, so that
\[
\mathcal{C\ell}(H_{V})=\mathcal{C\ell}^{(0)}(H_{V})\oplus\mathcal{C\ell}^{(1)}(H_{V})
\]
and the same notation of the exterior algebra is used for grade involution,
reversion, and conjugation in $\mathcal{C\ell}(H_{V})$, which, obviously,
satisfy
\[
(\mathbf{uv})\hat{\ }=\hat{\mathbf{u}}\hat{\mathbf{v}},\qquad(\mathbf{uv})
\tilde{\ }=\tilde{\mathbf{v}}\tilde{\mathbf{u}},\qquad(\mathbf{uv})\bar
{\ }=\bar{\mathbf{v}}\bar{\mathbf{u}}.
\]

For vecfors $\mathbf{x},\mathbf{y}\in H_{V}$, we have the relation
\begin{equation}
\mathbf{xy}+\mathbf{yx}=2\langle\mathbf{x},\mathbf{y}\rangle. \label{a60}
\end{equation}
In particular, just as with multivectors, in the Clifford algebra the norm map
$N:\mathcal{C\ell}(H_{V})\rightarrow\mathcal{C\ell}(H_{V})$ can be defined as
\begin{equation}
N\left(\mathbf{x}\right)=\mathbf{x}\bar{\mathbf{x}}, \label{a600}
\end{equation}
the square of the volume $2n$-vecfor $\mathbf{\sigma}$ satisfies
\[
\mathbf{\sigma}^{2}=1,
\]
and some properties of the Clifford product that will be useful in this work
are as follows: for a homogeneous multivecfor
$\mathbf{u}\in{\textstyle\bigwedge^{r}}H_{V}$, the grade involution is
$\widehat{\mathbf{u}}=\left(-1\right)^{r}\mathbf{u}$; in particular,
$\widehat{\mathbf{\sigma}}=\left(-1\right)^{2n}\mathbf{\sigma}=\mathbf{\sigma}$.
On the other hand,
\begin{equation}
\mathbf{u}\lrcorner\mathbf{\sigma}=\mathbf{u\sigma},\quad\mathbf{\sigma}\llcorner
\mathbf{u}=\mathbf{\sigma u} \label{a61}
\end{equation}
and if $\mathbf{x}\in H_{V}$,
\[
\mathbf{x}\lrcorner\mathbf{u}=\frac{1}{2}\left(\mathbf{xu}-\widehat
{\mathbf{u}}\mathbf{x}\right),\quad\mathbf{u}\lrcorner\mathbf{x}
=\frac{1}{2}\left(\mathbf{ux}-\mathbf{x}\widehat{\mathbf{u}}\right)
\]
\[
\mathbf{x}\wedge\mathbf{u}=\frac{1}{2}\left(\mathbf{xu}+\widehat{\mathbf{u}}
\mathbf{x}\right),\quad\mathbf{u}\wedge\mathbf{x}=\frac{1}{2}\left(
\mathbf{ux}+\mathbf{x}\widehat{\mathbf{u}}\right).
\]
Also note that
\begin{equation}
\mathbf{x\sigma}=-\widehat{\mathbf{\sigma}}\mathbf{x}=-\mathbf{\sigma x}. \label{a62}
\end{equation}
The following results on isomorphisms can be found in~\cite{qro7}.

\begin{Proposition}
There is the following natural isomorphism
\[
\mathcal{C\ell}(H_{V})\simeq\operatorname*{End}({\displaystyle\bigwedge}V).
\]
In addition, $b$ being a non-degenerate symmetric bilinear form on $V$, it
holds also
\[
\mathcal{C\ell}(H_{V})\simeq\mathcal{C\ell}(H_{bV})\simeq\mathcal{C\ell}
(V,b)\ \widehat{\otimes}\ \mathcal{C\ell}(V,-b),
\]
where $\widehat{\otimes}$ denotes the graded tensor product.
\end{Proposition}

\begin{Corollary}
The even and odd subspaces of the hyperbolic Clifford algebra are
\[
\mathcal{C\ell}^{(0)}(H_{V})\simeq\operatorname{End}(
{\displaystyle\bigwedge\nolimits^{(0)}}V)\oplus\operatorname{End}(
{\displaystyle\bigwedge\nolimits^{(1)}}V)
\]
and
\[
\mathcal{C\ell}^{(1)}(H_{V})\simeq L({\displaystyle\bigwedge\nolimits^{(0)}}V,
{\displaystyle\bigwedge\nolimits^{(1)}}V)\oplus L(
{\displaystyle\bigwedge\nolimits^{(1)}}V,{\displaystyle\bigwedge\nolimits^{(0)}}V),
\]
where $L(V,W)$ denotes the space of the linear mappings from $V$ to $W$ and
${\textstyle\bigwedge\nolimits^{(0)}}V$ and ${\textstyle\bigwedge\nolimits^{(1)}}V$
denote, respectively, the spaces of the even and of the odd elements of
${\textstyle\bigwedge}V$.
\end{Corollary}

\begin{table}[H]
\centering
\caption{Comparison between the standard Clifford algebra $\mathcal{C}\ell(V,g)$ and the Hyperbolic Clifford algebra $\mathcal{C}\ell(H_V)$.}
\label{tab:comparison}
\small
\begin{tabularx}{\textwidth}{>{\raggedright\arraybackslash}X>{\raggedright\arraybackslash}X>{\raggedright\arraybackslash}X}
\toprule
\textbf{Property} & $\boldsymbol{\mathcal{C}\ell(V,g)}$ & $\boldsymbol{\mathcal{C}\ell(H_V)}$ \\
\midrule
Underlying space & $(V, g)$, $\dim V = n$ & $H_V = V \oplus V^*$, $\dim H_V = 2n$ \\
Bilinear form & Non-degenerate, signature $(p,q)$ & Neutral, signature $(n,n)$ \\
Elements & Multivectors & Multivecfors ($v \oplus \omega$, $v\in V$, $\omega\in V^*$) \\
Algebra & $\mathcal{C}\ell(V,g)$ & $\mathcal{C}\ell(H_V) \simeq \operatorname{End}(\bigwedge V)$ \\
Relation & Subalgebra of $\mathcal{C}\ell(H_V)$ & ``Mother algebra'': contains $\mathcal{C}\ell(V,b)$ and $\mathcal{C}\ell(V,-b)$ \\
Isotropic subspaces & Depend on signature of $g$ & $V$ and $V^*$ are maximal totally isotropic \\
Superfield applications & Limited & Natural framework for superfields \\
\bottomrule
\end{tabularx}
\end{table}

\section{Groups Associated with $\mathcal{C\ell}(H_{V})$}\label{sec3}

Next, we will describe the groups and transformations involved in the spinor
structure of the Hyperbolic Clifford algebra. Just as in the Clifford algebra
of multivectors or multiforms, the associated groups are the orthogonal group,
the Clifford--Lipschitz group and the Pin and Spin groups, which in the case of
$\mathcal{C\ell}(H_{V})$ are appropriately defined.

\subsection{Orthogonal Transformations and Reflections in $H_{V}$}

Let $g$ be a symmetric bilinear form defined on a vector space $V$. A map
$T:V\rightarrow V$ is an isometry or an orthogonal transformation if
\[
g\left(T\left(v\right),T\left(u\right)\right)=g\left(v,u\right),
\qquad\forall v,u\in V;
\]
it immediately follows that $\left(\det T\right)^{2}=1$, i.e., $\det T=\pm1$.

Orthogonal transformations with $\det T=1$ are called rotations and those with
$\det T=-1$ are called reflections. The orthogonal group on the vector space
$H_{V}=V\oplus V^{\ast}$ will be denoted by
\[
O_{H_{V}}=\left\{ \text{Isometries } T:H_{V}\rightarrow H_{V}\right\}.
\]
The subgroup of $O_{H_{V}}$ with $\det T=1$ is called the special orthogonal
group and will be denoted by $SO_{H_{V}}$.

It is very important for the development of a theory in $\mathcal{C\ell}(H_{V})$
to specify the reflections. We know that the spaces $V$ and $V^{\ast}$
are maximal totally isotropic subspaces of $H_{V}$. Then these spaces can
be identified in $H_{V}$ under the image of the inclusions
$i_{\ast}:V\rightarrow H_{V}$, $x_{\ast}\mapsto i_{\ast}x_{\ast}=x_{\ast}\oplus0\equiv x_{\ast}$
and $i^{\ast}:V^{\ast}\rightarrow H_{V}$, $x^{\ast}\mapsto i^{\ast}x^{\ast}=0\oplus x^{\ast}\equiv x^{\ast}$,
so given a linear map $\phi\in\operatorname*{End}V$ (or similarly
$\phi^{\ast}\in\operatorname*{End}V^{\ast}$), it can be seen in
$\operatorname*{End}H_{V}$ under the inclusion $i_{\ast}$ (or under the
inclusion $i^{\ast}$, respectively). Thus, every linear map
$\phi\in\operatorname*{End}V$ (or, equivalently,
$\phi^{\ast}\in\operatorname*{End}V^{\ast}$) induces a linear map
$I\left(\phi\right)\in\operatorname*{End}H_{V}$, called the isotropic extension
of $\phi$ (see~\cite{qro7}), defined by
\begin{equation}
I\left(\phi\right)=\phi\oplus\phi^{\ast}. \label{a7}
\end{equation}

In this sense, every vecfor $\mathbf{x}\in H_{V}$ can be associated with a
reflection $S_{\mathbf{x}}:V\rightarrow V$ and a reflection
$S^{\mathbf{x}}:V^{\ast}\rightarrow V^{\ast}$ (here
$V,V^{\ast}\hookrightarrow H_{V}$ under the inclusions $i_{\ast}$ and
$i^{\ast}$, respectively) defined, respectively, by
\begin{equation}
S_{\mathbf{x}}y_{\ast}=y_{\ast}-2\frac{x^{\ast}\left(y_{\ast}\right)}
{x^{\ast}\left(x_{\ast}\right)}x_{\ast}\quad\text{and}\quad S^{\mathbf{x}}
y^{\ast}=y^{\ast}-2\frac{y^{\ast}\left(x_{\ast}\right)}{x^{\ast}\left(
x_{\ast}\right)}x^{\ast}. \label{a8}
\end{equation}

So, we can define the isotropic extension for the space $H_{V}$,
$I\left(S_{\mathbf{x}}\right)=\mathbf{S}_{\mathbf{x}}:H_{V}\rightarrow H_{V}$, given by
\[
\mathbf{S}_{\mathbf{x}}=S_{\mathbf{x}}\oplus S^{\mathbf{x}},
\]
or explicitly, for $\mathbf{y}\in H_{V}$, we have
\begin{equation}
\mathbf{S}_{\mathbf{x}}\mathbf{y}=\mathbf{y}-2\dfrac{\left\langle \mathbf{x},\mathbf{y}
\right\rangle }{\left\langle \mathbf{x},\mathbf{x}\right\rangle }\mathbf{x}. \label{a9}
\end{equation}
We immediately notice that
\begin{equation}
\mathbf{S}_{\mathbf{x}}\mathbf{x}=\mathbf{x}-2\dfrac{\left\langle \mathbf{x},\mathbf{x}
\right\rangle }{\left\langle \mathbf{x},\mathbf{x}\right\rangle }\mathbf{x}=-\mathbf{x}.
\label{a10}
\end{equation}
We also easily observe that $\mathbf{S}_{\mathbf{x}}$ is an orthogonal
map with respect to the bilinear form $\left\langle ,\right\rangle$
of $H_{V}$, i.e.,
\[
\left\langle \mathbf{S}_{\mathbf{x}}\mathbf{y},\mathbf{S}_{\mathbf{x}}\mathbf{z}
\right\rangle =\left\langle \mathbf{y},\mathbf{z}\right\rangle,
\]
for all $\mathbf{y},\mathbf{z}\in H_{V}$.

Furthermore, we can show that $\det\mathbf{S}_{\mathbf{x}}=-1$. In fact, we
already know that $\mathbf{S}_{\mathbf{x}}\mathbf{x}=-\mathbf{x}$; then consider
the volume element $\mathbf{\sigma}=\mathbf{\sigma}_{1}\wedge\ldots\wedge\mathbf{\sigma}_{2n}$,
where $\left\{ \mathbf{\sigma}_{1},\ldots,\mathbf{\sigma}_{2n}\right\}$ is an
orthonormal basis of $H_{V}$, and define the action of $\mathbf{S}_{x}$ on an
element $\mathbf{z}=z_{1}\wedge\cdots\wedge z_{p}\in\bigwedge^{p}H_{V}$ by
$\mathbf{S}_{x}\left(\mathbf{z}\right)=\mathbf{S}_{x}\left(z_{1}\right)\mathbf{S}_{x}\left(z_{2}\right)
\cdots\mathbf{S}_{x}\left(z_{p}\right)$. On the other hand, we have the relation
(see, for example,~\cite{Doran})
$\mathbf{S}_{\mathbf{x}}\left(\mathbf{\sigma}\right)=\left(\det\mathbf{S}_{\mathbf{x}}\right)\mathbf{\sigma}$,
and using the properties (\ref{a61}) and (\ref{a62}), we have
\[
\begin{array}[c]{lll}
\left(\det\mathbf{S}_{\mathbf{x}}\right)\mathbf{\sigma} & = &
\mathbf{S}_{\mathbf{x}}\left(\mathbf{\sigma}\right)=\mathbf{S}_{\mathbf{x}}
\left(\mathbf{\sigma}_{1}\mathbf{\sigma}_{2}\ldots\mathbf{\sigma}_{2n}\right)\\
& = & \mathbf{S}_{\mathbf{x}}\left(\mathbf{\sigma}_{1}\right)
\mathbf{S}_{\mathbf{x}}\left(\mathbf{\sigma}_{2}\right)\cdots
\mathbf{S}_{\mathbf{x}}\left(\mathbf{\sigma}_{2n}\right)\\
& = & \left(-1\right)^{2n}\mathbf{x\sigma x}^{-1}=-\left(-1\right)
^{2n}\mathbf{\sigma xx}^{-1}=-\mathbf{\sigma},
\end{array}
\]
whence we obtain $\det\mathbf{S}_{\mathbf{x}}=-1$, from which we conclude that
$\mathbf{S}_{\mathbf{x}}$ is a reflection.

On the other hand, in terms of $\mathcal{C\ell}(H_{V})$, from (\ref{a60}), we
have for vecfors $\mathbf{x},\mathbf{y}\in H_{V}$ the relationship
\begin{equation}
\mathbf{xy}+\mathbf{yx}=2\langle\mathbf{x},\mathbf{y}\rangle, \label{a11}
\end{equation}
from which $\mathbf{x}^{2}=\langle\mathbf{x},\mathbf{x}\rangle$, so the
vecfor $\mathbf{x}/\langle\mathbf{x},\mathbf{x}\rangle$ can be interpreted as
the inverse of $\mathbf{x}$, since
\begin{equation}
\mathbf{x}\frac{\mathbf{x}}{\langle\mathbf{x},\mathbf{x}\rangle}
=\frac{\mathbf{x}}{\langle\mathbf{x},\mathbf{x}\rangle}\mathbf{x}
=\frac{\mathbf{x}^{2}}{\langle\mathbf{x},\mathbf{x}\rangle}=\mathbf{1}. \label{a12}
\end{equation}
So from (\ref{a9}) and (\ref{a11}) we can write
\begin{equation}
\begin{array}[c]{lll}
\mathbf{S}_{\mathbf{x}}\mathbf{y} & = & \mathbf{y}-\left[\dfrac
{\mathbf{xy}+\mathbf{yx}}{\langle\mathbf{x},\mathbf{x}\rangle}\right]
\mathbf{x}\bigskip\\
& = & \mathbf{y}-\left[\mathbf{xy}+\mathbf{yx}\right]\mathbf{x}^{-1}\bigskip\\
& = & -\mathbf{xyx}^{-1}=\widehat{\mathbf{x}}\mathbf{yx}^{-1}.
\end{array}
\label{a13}
\end{equation}

In conclusion, a reflection in the hyperplane orthogonal to the vecfor
$\mathbf{x}$, in terms of $\mathcal{C\ell}(H_{V})$, satisfies the
identity\footnote{In~\cite{qro7} a matrix representation of the reflection in
the orthonormal basis of a hyperbolic space can be found.}
\begin{equation}
\mathbf{S}_{\mathbf{x}}\mathbf{y}=\widehat{\mathbf{x}}\mathbf{yx}^{-1},
\qquad\forall\mathbf{x},\mathbf{y}\in H_{V}. \label{a14}
\end{equation}

\subsection{The Clifford--Lipschitz Group Associated with $\mathcal{C\ell}(H_{V})$}

To define the Clifford--Lipschitz group associated with $\mathcal{C\ell}(H_{V})$,
we will try to preserve the notation as in the case of the multivector
algebra $\mathcal{C\ell}(V)$. Then, taking the vector space $V=\mathbb{R}^{p,q}$,
that is, $H_{V}=H_{\mathbb{R}^{p,q}}=\mathbb{R}^{p,q}\oplus\left(\mathbb{R}^{p,q}\right)^{\ast}$,
and to simplify the notation, instead of writing $H_{\mathbb{R}^{p,q}}$ we will
write $H_{p,q}$ and denote $\mathcal{C\ell}(H_{\mathbb{R}^{p,q}})$ as
$\mathcal{C\ell}_{H_{p,q}}$. Following this same reasoning, the orthogonal group
and special orthogonal group will be denoted by $O_{H_{p,q}}$ and
$SO_{H_{p,q}}$, respectively, for $V=\mathbb{R}^{p,q}$.

As in the multivector algebra, the most natural group of $\mathcal{C\ell}(H_{V})$
is the group of invertible elements of $\mathcal{C\ell}_{H_{p,q}}$, denoted by
\[
\mathcal{C\ell}_{H_{p,q}}^{\ast}=\left\{ \mathbf{x}\in\mathcal{C\ell}
_{H_{p,q}}:\exists\mathbf{x}^{-1}\right\}.
\]
This group acts naturally on $\mathcal{C\ell}_{H_{p,q}}$ as an algebra of
homomorphisms via its adjoint representation $Ad$ or twisted adjoint
representation $\widehat{A}d$, respectively:
\[
Ad:\mathcal{C\ell}_{H_{p,q}}^{\ast}\rightarrow\operatorname*{Aut}\left(
\mathcal{C\ell}_{H_{p,q}}\right);\ \mathbf{x}\longmapsto
Ad_{\mathbf{x}},\ \text{with}\ Ad_{\mathbf{x}}\left(\mathbf{u}\right)
=\mathbf{xux}^{-1}
\]
\[
\widehat{A}d:\mathcal{C\ell}_{H_{p,q}}^{\ast}\rightarrow\operatorname*{Aut}
\left(\mathcal{C\ell}_{H_{p,q}}\right);\ \mathbf{x}\longmapsto\widehat
{A}d_{\mathbf{x}},\ \text{with}\ \widehat{A}d_{\mathbf{x}}\left(
\mathbf{u}\right)=\widehat{\mathbf{x}}\mathbf{ux}^{-1}.
\]

To define a cover of the group of orthogonal transformations of $H_{p,q}$, one
can use its twisted adjoint action on $H_{p,q}$,
$\widehat{A}d_{\mathbf{x}}\left(u\right)=\widehat{\mathbf{x}}\mathbf{ux}^{-1}$,
since from (\ref{a14}) it is a reflection in the hyperplane orthogonal to the
vecfor $\mathbf{x}$, and then it automatically preserves the quadratic form in
$H_{p,q}$, i.e., $\widehat{A}d_{\mathbf{x}}\in O_{H_{p,q}}$. In what follows, we
will denote the twisted adjoint action $\widehat{A}d_{\mathbf{x}}$ by
$\rho_{\mathbf{x}}$. Thus, using $\rho_{\mathbf{x}}$ we can define the
Clifford--Lipschitz group for $H_{p,q}$ as the subgroup of
$\mathcal{C\ell}_{H_{p,q}}^{\ast}$ defined by
\begin{equation}
\Gamma_{H_{p,q}}=\left\{ \mathbf{x}\in\mathcal{C\ell}_{H_{p,q}}^{\ast}:
\alpha\left(\mathbf{x}\right)\mathbf{vx}^{-1}\in H_{p,q},\qquad
\forall\mathbf{v}\in H_{p,q}\right\}, \label{CL0}
\end{equation}
where $\alpha\left(\mathbf{x}\right)\equiv\widehat{\mathbf{x}}$ is the
grade involution, a notation that we will also adopt in what follows.

Furthermore, the twisted adjoint action $\rho$ of the group $\Gamma_{H_{p,q}}$
on the hyperbolic vector space $H_{p,q}$ is defined by
\begin{equation}
\rho:\Gamma_{H_{p,q}}\rightarrow\operatorname*{Aut}\left(H_{p,q}\right),
\qquad\rho_{\mathbf{x}}\mathbf{v}=\alpha\left(\mathbf{x}\right)
\mathbf{vx}^{-1},\quad\forall\mathbf{v}\in H_{p,q}. \label{CL1}
\end{equation}

It is easy to show that $\rho_{\mathbf{x}}$ is a linear action on $H_{p,q}$
and that $\Gamma_{H_{p,q}}$ is a group. Furthermore, note that
$\rho_{\mathbf{x}}$ is an isomorphism on $H_{p,q}$ for all
$\mathbf{x}\in\Gamma_{H_{p,q}}$. Indeed, first note that $\rho_{\mathbf{x}}$ is
injective: suppose that $\rho_{\mathbf{x}}\left(u\right)=0$, so we have that
$\alpha\left(\mathbf{x}\right)u\mathbf{x}^{-1}=0$; then, multiplying on
the right by $\mathbf{x}$ and on the left by $\alpha\left(\mathbf{x}^{-1}\right)$
we obtain that $u=0$, that is, $\ker\rho_{\mathbf{x}}=\left\{0\right\}$, and
since $\dim H_{p,q}$ is finite, $\rho_{\mathbf{x}}:H_{p,q}\rightarrow H_{p,q}$
is an isomorphism $\forall\mathbf{x}\in\Gamma_{H_{p,q}}$.

\begin{Theorem}
The map $\rho:\Gamma_{H_{p,q}}\rightarrow O_{H_{p,q}}$ is a surjective
homomorphism whose kernel is the multiplicative group $\mathbb{R}^{\ast}\mathbf{1}$
of the nonzero scalar multiples of the unit of $\mathcal{C\ell}_{H_{p,q}}$. The
restriction of $N\left(\mathbf{x}\right)$ to $\Gamma_{H_{p,q}}$ is a nonzero scalar.
\end{Theorem}

\begin{proof}
Let us first note that if $\mathbf{x}\in\ker\rho$, by definition
$\rho_{\mathbf{x}}v=v$, that is,
$\rho_{\mathbf{x}}v=\alpha\left(\mathbf{x}\right)v\mathbf{x}^{-1}=v$; then
$\alpha\left(\mathbf{x}\right)v=v\mathbf{x}$ $\forall v\in H_{p,q}$.

Since $\mathbf{x}\in\Gamma_{H_{p,q}}$, in particular,
$\mathbf{x}\in\mathcal{C\ell}_{H_{p,q}}=\mathcal{C\ell}_{H_{p,q}}^{(0)}\oplus\mathcal{C\ell}_{H_{p,q}}^{(1)}$,
we can separate the even part and the odd part of $\mathbf{x}$, that is,
\begin{equation}
\mathbf{x}=x_{0}+x_{1}, \label{CL2}
\end{equation}
where $x_{0}$ is even and $x_{1}$ is odd; thus
$\alpha\left(\mathbf{x}\right)=x_{0}-x_{1}$. Then,
\[
\alpha\left(\mathbf{x}\right)v=v\left(x_{0}+x_{1}\right)\qquad\forall v\in H_{p,q}.
\]

Then, setting $x_{0}=a_{0}+a_{1}\sigma_{1}$ and $x_{1}=b_{0}\sigma_{1}+b_{1}$,
where $a_{0},b_{0}$ are even and $a_{1},b_{1}$ are odd and all are elements of
$\mathcal{C\ell}_{H_{p,q}}$ independent of $\sigma_{1}$, substituting into the
condition $\alpha(\mathbf{x})v = v\mathbf{x}$ for all $v \in H_{p,q}$ and taking
$v = \sigma_1$, one obtains
\[
(x_0 - x_1)\sigma_1 = \sigma_1(x_0 + x_1).
\]
Using $\sigma_1^2 = \langle \sigma_1, \sigma_1\rangle \in \mathbb{R}^*$ and the
anticommutation relations of the basis vectors, this forces $a_1 = 0$ and
$b_0 = 0$, i.e., $x_0$ and $x_1$ are independent of $\sigma_1$. Following the
same reasoning for each basis element $\sigma_k$ with $k=1,\ldots,2n$
(see~\cite{Todorov,Vaz} for full details), the argument is valid for all
elements of the basis $\left\{ \sigma_{1},\sigma_{2},\ldots,\sigma_{2n}\right\}$
of $\mathcal{C\ell}_{H_{p,q}}$, and therefore, if $\mathbf{x}\in\ker\rho$ does
not depend on any element of the basis, then $\mathbf{x}$ commutes with every
basis element of $\mathcal{C\ell}_{H_{p,q}}$. Since the center of
$\mathcal{C\ell}_{H_{p,q}}$ consists only of real scalar multiples of the
identity (for $p+q \geq 1$), $\mathbf{x}$ is a real number multiplied by the
unit of $\mathcal{C\ell}_{H_{p,q}}$.

On the other hand, since $\alpha\left(\overline{u}\right)=u$ $\forall u\in H_{V}$,
if $\mathbf{x}\in\Gamma_{H_{p,q}}$ we have
$\alpha\left(\mathbf{x}\right)v\mathbf{x}^{-1}\in H_{V}$, which implies that
\[
\alpha\left(\mathbf{x}\right)v\mathbf{x}^{-1}=\alpha\left(\overline
{\alpha\left(\mathbf{x}\right)v\mathbf{x}^{-1}}\right)=\alpha\left(
\overline{\mathbf{x}^{-1}}\overline{v}\overline{\alpha\left(\mathbf{x}
\right)}\right)=\alpha\left(\overline{\mathbf{x}}^{-1}\right)v\overline{\mathbf{x}}.
\]
Thus, from the previous equality we have
\[
\begin{array}[c]{ll}
\alpha\left(N\left(\mathbf{x}\right)^{-1}\right)vN\left(\mathbf{x}\right)
& =\alpha\left(\left(\mathbf{x}\overline{\mathbf{x}}\right)^{-1}\right)
v\mathbf{x}\overline{\mathbf{x}}=\alpha\left(\overline{\mathbf{x}}^{-1}\right)
\alpha\left(\mathbf{x}^{-1}\right)v\mathbf{x}\overline{\mathbf{x}}\bigskip\\
& =\alpha\left(\mathbf{x}\right)\alpha\left(\mathbf{x}^{-1}\right)
v\mathbf{xx}^{-1}=v,
\end{array}
\]
from which, for $\mathbf{x}\in\Gamma_{H_{p,q}}$, we have that
$N\left(\mathbf{x}\right)\in\mathbb{R}^{\ast}\mathbf{1}$.
\end{proof}

\begin{Remark}
The surjective homomorphism $\rho : \Gamma_{H_{p,q}} \to \operatorname{O}_{H_{p,q}}$
with kernel $\mathbb{R}^*$ parallels the classical construction for standard
Clifford algebras. The fundamental difference lies in the structure: while
standard Clifford--Lipschitz groups act on $n$-dimensional vector spaces, the
hyperbolic version acts on $2n$-dimensional hyperbolic structures $H_{p,q}$.
This allows the simultaneous representation of both multivectors
($\bigwedge V$) and multiforms ($\bigwedge V^*$), providing a unified algebraic
framework absent in classical theory.
\end{Remark}

\subsection{The Pin and Spin Groups of $\mathcal{C\ell}_{H_{p,q}}$}

The Pin and Spin groups are the most ``reduced'' subgroups of $\Gamma_{H_{p,q}}$
to describe orthogonal transformations. These groups are defined by
restricting the group $\Gamma_{H_{p,q}}$ to those elements that have norm
$\pm1$. That is, the group $\operatorname{Pin}_{H_{p,q}}$ is defined as
\[
\operatorname*{Pin}\nolimits_{H_{p,q}}=\left\{ a\in\Gamma_{H_{p,q}}:N\left(
a\right)=\pm1\right\};
\]
then, restricting the domain of the twisted adjoint action $\rho$ to the group
$\operatorname*{Pin}\nolimits_{H_{p,q}}$,
\[
\rho:\operatorname*{Pin}\nolimits_{H_{p,q}}\rightarrow O_{H_{p,q}},
\]
we obtain that
\[
\left.\ker\rho\right\vert_{\operatorname*{Pin}\nolimits_{H_{p,q}}}=\ker
\rho\cap\ker N=\left\{\pm1\right\}=\mathbb{Z}_{2}.
\]
The group $\operatorname*{Spin}_{H_{p,q}}$ is the subgroup of the group
$\Gamma_{H_{p,q}}^{+}=\Gamma_{H_{p,q}}\cap\mathcal{C\ell}_{H_{p,q}}^{+}$, defined as
\[
\operatorname*{Spin}\nolimits_{H_{p,q}}=\left\{ a\in\Gamma_{H_{p,q}}^{+}:N\left(a\right)=\pm1\right\}.
\]
So, the twisted adjoint action $\rho$ restricted to the group
$\operatorname*{Spin}\nolimits_{H_{p,q}}$ has its image in $SO_{H_{p,q}}$, that is,
\begin{equation}
\rho:\operatorname*{Spin}\nolimits_{H_{p,q}}\rightarrow SO_{H_{p,q}}, \label{CL3}
\end{equation}
and also $\left.\ker\rho\right\vert_{\operatorname*{Spin}_{H_{p,q}}}=\left\{\pm1\right\}=\mathbb{Z}_{2}$.

The above results can be summarized in the following theorem, similar to the
case of the multivector Clifford algebra $\mathcal{C\ell}(V)$.

\begin{Theorem}
Let $\rho:\Gamma_{H_{p,q}}\rightarrow O_{H_{p,q}}$ be the surjective
homomorphism studied previously; then we have the following relations
depending on the domain of $\rho$:
\begin{enumerate}
\item[(a)] $\left.\rho\right\vert_{\operatorname*{Pin}\nolimits_{H_{p,q}}}:
\operatorname*{Pin}\nolimits_{H_{p,q}}\rightarrow O_{H_{p,q}}$ is surjective
with kernel $\mathbb{Z}_{2}$.

\item[(b)] $\left.\rho\right\vert_{\operatorname*{Spin}_{H_{p,q}}}:
\operatorname*{Spin}_{H_{p,q}}\rightarrow SO_{H_{p,q}}$ is surjective with
kernel $\mathbb{Z}_{2}$.

\item[(c)] $\left.\rho\right\vert_{\operatorname*{Spin}_{H_{p,q}}^{e}}:
\operatorname*{Spin}_{H_{p,q}}^{e}\rightarrow SO_{H_{p,q}}^{e}$ is
surjective with kernel $\mathbb{Z}_{2}$.

\item[(d)]
\[
O_{H_{p,q}}=\frac{\operatorname*{Pin}\nolimits_{H_{p,q}}}{\mathbb{Z}_{2}};\
SO_{H_{p,q}}=\frac{\operatorname*{Spin}_{H_{p,q}}}{\mathbb{Z}_{2}};\
SO_{H_{p,q}}^{e}=\frac{\operatorname*{Spin}_{H_{p,q}}^{e}}{\mathbb{Z}_{2}}.
\]
\end{enumerate}
\end{Theorem}

\begin{Remark}
The definitions of $\operatorname{Pin}_{H_{p,q}}$ and
$\operatorname{Spin}_{H_{p,q}}$ via restriction to elements of norm $\pm 1$
follow the standard construction. However, the hyperbolic framework introduces
a crucial new feature: both $\operatorname{SO}_{H_{p,q}}$ (the special
orthogonal group) and its oriented version $\operatorname{SO}^e_{H_{p,q}}$
admit double covers via the respective Spin groups. This is essential for
spinor constructions or representations on hyperbolic Lorentzian manifolds and
for developing consistent spinor structures on spacetime with simultaneous
treatment of field components and their duals.
\end{Remark}

\section{Spinor Bundle Structure of the Hyperbolic Clifford Algebra}\label{sec4}

To introduce the spinor bundle on $\mathcal{C\ell}\left(H_{V}\right)$, we must
take into account that in the case of hyperbolic space, two vector bundles are
involved, the tangent bundle and the cotangent bundle, on the same manifold
$M$; let us say $\left(E_{1},M,\pi_{1},G,V_{1}\right)$ and
$\left(E_{2},M,\pi_{2},G,V_{2}\right)$. The Whitney sum of $E_{1}$ and $E_{2}$ is
the vector bundle with total space $E_{1}\oplus E_{2}$, which is given by
\[
E_{1}\oplus E_{2}=\left\{ \left(p_{1},p_{2}\right)\in E_{1}\times
E_{2}:\pi_{1}\left(p_{1}\right)=\pi_{2}\left(p_{2}\right)\right\},
\]
with projection $\Pi:E_{1}\oplus E_{2}\rightarrow M$ given by
\begin{equation}
\Pi\left(\left(p_{1},p_{2}\right)\right)=\pi_{1}\left(p_{1}\right)
=\pi_{2}\left(p_{2}\right). \label{FH1}
\end{equation}

Then, the fiber of $E_{1}\oplus E_{2}$ corresponding to $x\in M$ is the direct
sum of the vector spaces $\pi_{1}^{-1}\left(x\right)\oplus\pi_{2}^{-1}\left(x\right)$.
To verify the local trivialization condition in $E_{1}\oplus E_{2}$, it suffices
to take trivializing functions $\left\{ \varphi_{\alpha}\right\}$ and
$\left\{ \psi_{\alpha}\right\}$ for $E_{1}$ and $E_{2}$, respectively,
subordinated to the same cover $\left(U_{\alpha}\right)_{\alpha\in I}$ of $M$,
whose induced transition functions are
$s_{\alpha\beta}=\varphi_{\alpha}\circ\varphi_{\beta}^{-1}$ and
$h_{\alpha\beta}=\psi_{\alpha}\circ\psi_{\beta}^{-1}$; then the trivializations
for $E_{1}\oplus E_{2}$ are given by
\begin{equation}
\left\{ \phi_{\alpha}=\left(\varphi_{\alpha}\times\psi_{\alpha}\right)\right\}, \label{FH2}
\end{equation}
to which correspond the transition functions
\begin{equation}
g_{\alpha\beta}=s_{\alpha\beta}\oplus h_{\alpha\beta}=\left[
\begin{array}[c]{cc}
s_{\alpha\beta} & 0\\
0 & h_{\alpha\beta}
\end{array}
\right], \label{FH10}
\end{equation}
recalling that $g_{\alpha\beta}:U_{\alpha}\cap U_{\beta}\rightarrow G$; from
(\ref{FH10}) we will have $G=Gl\left(2n,\mathbb{R}\right)$. For details
on the Whitney sum, see, e.g.,~\cite{Osborn}.

A very useful and fundamental property of the Whitney sum for the next topic
is that if $\mathcal{F}_{1}$ and $\mathcal{F}_{2}$ are trivial vector bundles
with vector spaces of dimension $n$ and $m$, respectively, then
$\mathcal{F}_{3}=\mathcal{F}_{1}\oplus\mathcal{F}_{2}$ is a trivial vector bundle
with vector space of dimension $n+m$, since there exists an isomorphism
\[
\chi:\mathcal{F}_{1}\oplus\mathcal{F}_{2}\rightarrow\mathcal{F}_{3}
\]
defined by $\chi\left(\left(x,u\right),\left(x,v\right)\right)=\left(x,u+v\right)$,
where $u+v\in V$, with $\dim V=n+m$. Then, to define a hyperbolic spinor
structure on $M$, we can use a known result due to Geroch~\cite{Geroch1,Geroch2},
which states that for a Lorentz manifold $(M,g)$, a spinor structure exists if
and only if the principal bundle $P_{SO_{1,3}^{e}}\left(M\right)$ is a trivial
bundle. Recall that a principal bundle is trivial if and only if it admits a
global section. Therefore, Geroch's result says that a spacetime admits a
spinor structure if and only if it admits a global Lorentz frame.

\begin{Remark}
It is natural to ask whether the existence of a hyperbolic spinor structure
imposes stricter topological constraints on $M$ than a standard spinor
structure. The existence of a hyperbolic spinor structure requires the
existence of a global section of the principal bundle
$\mathbf{P}_{SO^e_{H_{p,q}}}(M)$, i.e., a global hyperbolic orthonormal frame.
Since the structure group $SO^e_{H_{p,q}}$ has a richer topology than
$SO^e_{1,3}$, the precise topological obstructions depend on the homotopy type
of $SO^e_{H_{p,q}}$. A detailed analysis of these constraints lies beyond the
scope of the present paper and is left as a direction for future research.
\end{Remark}

\subsection{Frame Bundle on $\mathcal{C\ell}\left(H_{V}\right)$}

Similar to the tangent bundle $TM$, we will say that the hyperbolic tangent
bundle $TM\oplus T^{\ast}M$ of a differentiable $n$-dimensional manifold $M$
is associated to a principal bundle
$F_{H}\left(M\right)=\cup_{x\in M}F_{x}M\oplus F_{x}^{\ast}M$, called the frame
bundle, where $F_{x}M\oplus F_{x}^{\ast}M$ is the set of frames at $x\in M$. The
structure group of $F_{H}\left(M\right)$ is $Gl\left(2n,\mathbb{R}\right)$
(here, following the notation of a principal fiber bundle, the bundle
$F_{H}\left(M\right)$ is formally written as
$\left(F_{H}M,M,\pi,Gl\left(2n,\mathbb{R}\right)\right)$). Let
$\left\{ x_{\alpha}^{1},x_{\alpha}^{2},\cdots,x_{\alpha}^{n}\right\}$ be the
coordinates associated with a local chart $\left(U_{\alpha},\varphi_{\alpha}\right)$
of the maximal atlas of $M$. Thus, the natural basis of $T_{x}M\oplus T_{x}^{\ast}M$
on $U_{\alpha}\subset M$ is given by
\begin{equation}
\left\{ \dfrac{\partial}{\partial x^{1}}\oplus dx^{1},\cdots,\dfrac{\partial
}{\partial x^{n}}\oplus dx^{n},\dfrac{-\partial}{\partial x^{1}}\oplus
dx^{1},\cdots,\dfrac{-\partial}{\partial x^{n}}\oplus dx^{n}\right\}, \label{FH11}
\end{equation}
where the $\dfrac{-\partial}{\partial x^{k}}\oplus dx^{k}$ are obtained from
$\dfrac{\partial}{\partial x^{k}}\oplus dx^{k}$ through the involution
\[
\dfrac{\partial}{\partial x^{k}}\oplus dx^{k}\rightarrow\dfrac{-\partial
}{\partial x^{k}}\oplus dx^{k};
\]
see (\ref{a2}).

To simplify the notation relating to the indices, we set
\begin{equation}
\mathbf{\gamma}_{k}=\dfrac{\partial}{\partial x^{k}}\oplus dx^{k}\quad
\text{and}\quad\mathbf{\gamma}_{n+k}=\dfrac{-\partial}{\partial x^{k}}\oplus dx^{k}
\label{FH011}
\end{equation}
with $k=1,\ldots,n$. Then, from (\ref{FH11}) we can write the natural basis of
$T_{x}M\oplus T_{x}^{\ast}M$ over $U_{\alpha}\subset M$ as
\[
\left\{ \mathbf{\gamma}_{1},\cdots,\mathbf{\gamma}_{n},\mathbf{\gamma}
_{n+1},\cdots,\mathbf{\gamma}_{2n}\right\}.
\]

\begin{Definition}
A frame at $T_{x}M\oplus T_{x}^{\ast}M$ is a set
\[
\Theta_{x}=\left\{ \left.z_{1}\right\vert_{x},\cdots,\left.
z_{n}\right\vert_{x},\left.z_{n+1}\right\vert_{x},\cdots,\left.
z_{2n}\right\vert_{x}\right\}
\]
of linearly independent vectors such that
\begin{equation}
\left.z_{i}\right\vert_{x}=F_{i}^{j}\left.\mathbf{\gamma}_{j}\right\vert_{x}, \label{FH12}
\end{equation}
where the matrix $\left(F_{i}^{j}\right)\in Gl\left(2n,\mathbb{R}\right)$.
\end{Definition}

\begin{Remark}
In the standard tangent bundle $TM$, a frame at $T_x M$ consists of a basis of
the $n$-dimensional tangent space. Here, we extend this concept to the
hyperbolic tangent-cotangent bundle $TM \oplus T^*M$, where frames incorporate
both contravariant (tangent) and covariant (cotangent) directions. This
symmetric treatment of primal and dual spaces is a defining feature of the
hyperbolic structure and is essential for theories in general relativity that
require simultaneous treatment of fields and their dual counterparts.
\end{Remark}

A local trivialization
$\phi_{\alpha}:\Pi^{-1}\left(U_{\alpha}\right)\rightarrow U_{\alpha}\times Gl\left(2n,\mathbb{R}\right)$
of $F_{H}\left(M\right)$ is defined by
\begin{equation}
\phi_{\alpha}\left(p\right)=\left(\Pi\left(p\right),\varkappa
_{\alpha}\left(p\right)\right)=\left(x,\Theta_{x}\right), \label{FH13}
\end{equation}
where $\Pi\left(p\right)=x$ and $\varkappa_{\alpha}\left(p\right)=\Theta_{x}$.
Here $p=\left(p_{1},p_{2}\right)$, and $\Pi$ and $\phi_{\alpha}$ are as in
(\ref{FH1}) and (\ref{FH2}), respectively.

The action of $G=\left(G_{i}^{j}\right)\in Gl\left(2n,\mathbb{R}\right)$ on a
frame $\Omega\in F_{H}\left(U\right)$ is given by
$\left(\Omega,G\right)\rightarrow\Omega G$, where the new frame
$\Omega G\in F_{H}\left(U\right)$ is defined by
$\phi_{\alpha}\left(\Omega G\right)=\left(\Pi\left(\Omega G\right),\varkappa_{\alpha}\left(\Omega G\right)\right)=\left(x,\Theta_{x}^{\prime}\right)$,
with $\Pi\left(\Omega G\right)=x$ and
\[
\varkappa_{\alpha}\left(\Omega G\right)=\Theta_{x}^{\prime}=\left\{
\left.z_{1}^{\prime}\right\vert_{x},\cdots,\left.z_{2n}^{\prime}\right\vert_{x}\right\}
\]
with
\begin{equation}
\left.z_{i}^{\prime}\right\vert_{x}=\left.z_{j}\right\vert_{x}G_{i}^{j}, \label{FH14}
\end{equation}
where $i,j=1,\ldots,2n$. Conversely, given the frames $\Theta_{x}$ and
$\Theta_{x}^{\prime}$ there exists $G=\left(G_{i}^{j}\right)\in Gl\left(2n,\mathbb{R}\right)$
such that (\ref{FH14}) is satisfied, which means that
$Gl\left(2n,\mathbb{R}\right)$ acts on $F_{H}\left(M\right)$ actively.

Let $\left\{ x^{\alpha}\right\}$ and $\left\{ x^{\prime\alpha}\right\}$
be the coordinates associated with the local charts
$\left\{ U_{\alpha},\varphi_{\alpha}\right\}$ and
$\left\{ U_{\alpha}^{\prime},\varphi_{\alpha}^{\prime}\right\}$, respectively,
of the maximal atlas of $M$. If $x\in U_{\alpha}\cap U_{\beta}$ we have
\[
\left.z_{i}\right\vert_{x}=F_{i}^{j}\left.\mathbf{\gamma}_{j}\right\vert
_{x}=F_{i}^{\prime j}\left.\mathbf{\gamma}_{j}^{\prime}\right\vert_{x},
\]
where $\left(F_{i}^{j}\right),\left(F_{i}^{\prime j}\right)\in Gl\left(2n,\mathbb{R}\right)$,
and note that the $\mathbf{\gamma}_{l}^{\prime}$ are defined as in
(\ref{FH011}) in the $\left\{ x^{\prime\alpha}\right\}$ coordinates, i.e.,
\[
\mathbf{\gamma}_{l}^{\prime}=\dfrac{\partial}{\partial x^{\prime l}}\oplus
dx^{\prime l}\quad\text{and}\quad\mathbf{\gamma}_{n+l}^{\prime}
=\dfrac{-\partial}{\partial x^{\prime l}}\oplus dx^{\prime l},
\]
with $l=1,\ldots,n$. Since
$F_{i}^{j}=F_{k}^{\prime j}\left.\left(\dfrac{\partial x^{k}}{\partial x^{\prime j}}\right)\right\vert_{x}$
we have that the transition functions are
\begin{equation}
g_{i}^{k}\left(x\right)=\left.\left(\frac{\partial x^{k}}{\partial
x^{\prime j}}\right)\right\vert_{x}\in Gl\left(2n,\mathbb{R}\right). \label{FH16}
\end{equation}
Recall that transition functions are continuous functions
\begin{equation}
g_{\alpha\beta}:U_{\alpha}\cap U_{\beta}\rightarrow G,\text{ where }
g_{\alpha\beta}\left(x\right)=\phi_{\alpha,x}\circ\phi_{\beta,x}^{-1}, \label{FH17}
\end{equation}
and $g_{\alpha\beta}$ is as in (\ref{FH10}), i.e.,
\begin{equation}
\begin{array}[c]{l}
g_{\alpha\beta}\left(x\right)=s_{\alpha\beta}\left(x\right)\oplus
h_{\alpha\beta}\left(x\right)\bigskip\\
=\left[
\begin{array}[c]{cc}
s_{\alpha\beta}\left(x\right) & 0\\
0 & h_{\alpha\beta}\left(x\right)
\end{array}
\right]=\left[
\begin{array}[c]{cc}
\varphi_{\alpha,x}\circ\varphi_{\beta,x}^{-1} & 0\\
0 & \psi_{\alpha,x}\circ\psi_{\beta,x}^{-1}
\end{array}
\right].
\end{array}
\label{FH18}
\end{equation}

Now making use of the metric field defined by
$\left\langle \mathbf{x},\mathbf{y}\right\rangle =x^{\ast}\left(y_{\ast}\right)+y^{\ast}\left(x_{\ast}\right)$,
where $\mathbf{x}=x_{\ast}\oplus x^{\ast},\mathbf{y}=y_{\ast}\oplus y^{\ast}\in H_{V}$,
defined in (\ref{a1}), we can introduce orthonormal frames in each
$T_{x}U\oplus T_{x}^{\ast}U$. That is, for each $T_{x}U\oplus T_{x}^{\ast}U$, we
can denote an orthonormal frame by
$\Theta_{x}=\left\{ \left.z_{1}\right\vert_{x},\cdots,\left.z_{n}\right\vert_{x},\left.z_{n+1}\right\vert_{x},\cdots,\left.z_{2n}\right\vert_{x}\right\}$,
where
\[
\left.z_{i}\right\vert_{x}=h_{i}^{j}\left.\gamma_{j}\right\vert_{x},
\]
with $\left(h_{i}^{j}\right)\in O_{H_{p,q}}$, and where $O_{H_{p,q}}$ is
the $2n$-dimensional real orthogonal group. In this case, the frame bundle is
said to have been reduced to the hyperbolic orthonormal frame bundle, which
will be denoted by $\mathbf{P}_{O_{H_{p,q}}}\left(M\right)$, or, when
$\left(h_{i}^{j}\right)\in SO_{H_{p,q}}$, the hyperbolic frame bundle will
be denoted by $\mathbf{P}_{SO_{H_{p,q}}}\left(M\right)$.

In an analogous way, we can define a hyperbolic principal bundle of oriented
orthonormal frames $\mathbf{P}_{SO_{H_{p,q}}^{e}}\left(M\right)$ on a
Lorentzian manifold modeling spacetime, and its covering bundle, called the
hyperbolic spin bundle $\mathbf{P}_{\operatorname{Spin}_{H_{p,q}}^{e}}\left(M\right)$.
Recall that the isomorphism defined in (\ref{a3}) allows us to choose an
arbitrary non-degenerate symmetric bilinear form $b$. Thus, in the case of a
hyperbolic space, a Lorentzian manifold is a pair $\left(M,g\right)$,
where $g\in\sec T^{2,0}M$ is a Lorentzian metric of signature $\left(1,3\right)$,
i.e., for all $x\in M$, $T_{x}M\oplus T_{x}^{\ast}M\simeq\mathbb{R}_{H}^{1,3}$,
where $\mathbb{R}_{H}^{1,3}$ is the hyperbolic Minkowski vector space. Then, the
hyperbolic Clifford algebra of $\mathbb{R}_{H}^{1,3}$ will be denoted by
$\mathbb{R}_{1,3}^{H}$. Most of the properties of the hyperbolic algebra
$\mathbb{R}_{1,3}^{H}$ are inherited from the $\mathbb{R}_{1,3}$ algebra and are
fundamental to this theory; for example, it is straightforward to show that
every automorphism of $\mathbb{R}_{1,3}^{H}$ is inner, and if we denote by
$\mathbb{R}_{1,3}^{\ast H}$ the group of invertible elements of
$\mathbb{R}_{1,3}^{H}$, then through the adjoint representation this group acts
on $\mathbb{R}_{1,3}^{H}$ as an algebra of automorphisms. In addition, the group
$\operatorname{Spin}_{H_{p,q}}^{e}$ has a natural extension in $\mathbb{R}_{1,3}^{H}$.

\subsection{Hyperbolic Clifford Bundle}

\begin{Definition}
The Hyperbolic Clifford bundle of the metric manifold $M$ is
\[
\mathcal{C\ell}(H\left(M\right))\equiv\mathcal{C\ell}(TM\oplus T^{\ast}M,\left\langle {}\right\rangle)
={\textstyle\bigcup\limits_{x\in M}}\mathcal{C\ell}(T_{x}M\oplus T_{x}^{\ast}M,\left\langle {}\right\rangle),
\]
where $\mathcal{C\ell}(T_{x}M\oplus T_{x}^{\ast}M,\left\langle {}\right\rangle)$
is the hyperbolic Clifford algebra of the hyperbolic structure
\[
\left(T_{x}M\oplus T_{x}^{\ast}M,\left\langle {}\right\rangle \right).
\]
\end{Definition}

The hyperbolic Clifford bundle $\mathcal{C\ell}(H\left(M\right))$ is a
vector bundle associated with the principal bundle of orthonormal frames
$\mathbf{P}_{SO_{H_{1,3}}^{e}}\left(M\right)$ associated to a Lorentzian
manifold, i.e.,
\[
\mathcal{C\ell}(H\left(M\right))=\mathbf{P}_{SO_{H_{1,3}}^{e}}\left(M\right)
\times_{\rho}\mathcal{C\ell}(T_{x}M\oplus T_{x}^{\ast}M,\left\langle {}\right\rangle)
\simeq\mathbf{P}_{SO_{H_{1,3}}^{e}}\left(M\right)\times_{\rho}\mathbb{R}_{1,3}^{H}.
\]

Indeed, considering the canonical projection
$\Pi_{H}:\mathcal{C\ell}(H\left(M\right))\rightarrow M$ and taking
$\left(U_{\alpha}\right)_{\alpha\in I}$ an open covering of $M$, we can
define the trivialization mappings
\[
\varphi_{\alpha}:\Pi_{H}^{-1}\left(U_{\alpha}\right)\rightarrow U_{\alpha}\times\mathbb{R}_{1,3}^{H},
\]
such that
$\varphi_{\alpha}\left(p\right)=\left(\Pi_{H}(p),\varphi_{\alpha,x}\left(p\right)\right)=\left(x,\varphi_{\alpha,x}\left(p\right)\right)$.
Thus, if $x\in U_{\alpha}\cap U_{\beta}$ and $p\in\Pi_{H}^{-1}\left(x\right)$, we have
\[
\varphi_{\alpha,x}\left(p\right)=h_{\alpha\beta}\left(x\right)\varphi_{\beta,x}\left(p\right),
\]
for $h_{\alpha\beta}\left(x\right)\in\operatorname*{Aut}\left(\mathbb{R}_{1,3}^{H}\right)$,
where $h_{\alpha\beta}:U_{\alpha}\cap U_{\beta}\rightarrow\operatorname*{Aut}\left(\mathbb{R}_{1,3}^{H}\right)$
are the transition mappings of $\mathcal{C\ell}(H\left(M\right))$. Then, as
every automorphism of $\mathbb{R}_{1,3}^{H}$ is inner, we have
\[
h_{\alpha\beta}\left(x\right)\varphi_{\beta,x}\left(p\right)
=g_{\alpha\beta}\left(x\right)\varphi_{\alpha,x}\left(p\right)g_{\alpha\beta}\left(x\right)^{-1},
\]
for some $g_{\alpha\beta}\left(x\right)\in\mathbb{R}_{1,3}^{\ast H}$.

On the other hand, taking into account the isomorphism (\ref{CL3}), we can
deduce that the structure group of the hyperbolic Clifford bundle
$\mathcal{C\ell}(H\left(M\right))$ is reducible from
$\operatorname*{Aut}\left(\mathbb{R}_{1,3}^{H}\right)$ to $SO_{H_{1,3}}^{e}$, and
the transition maps for $\mathcal{C\ell}(H\left(M\right))$ can be taken from
$\mathbf{P}_{SO_{H_{1,3}}^{e}}\left(M\right)$. Then, the hyperbolic
Clifford bundle is a vector bundle associated to the principal bundle
$\mathbf{P}_{SO_{H_{1,3}}^{e}}\left(M\right)$, i.e.,
\[
\mathcal{C\ell}(H\left(M\right))=\mathbf{P}_{SO_{H_{1,3}}^{e}}\left(M\right)\times_{\rho}\mathbb{R}_{1,3}^{H}.
\]

Details on the construction of a vector bundle in the field of theoretical
physics can be found, e.g., in~\cite{Lawson,Rodrigues1}. Finally, we can
define a hyperbolic spinor structure on $M$, similarly to the case when the
spinor structure is given by a spin frame bundle; see, e.g.,~\cite{Rodrigues1},
from which we adopt the notation.

\begin{Definition}
A hyperbolic spinor structure on $M$ consists of a hyperbolic principal fiber
bundle with group $\operatorname{Spin}_{H_{1,3}}^{e}$,
\[
\Pi_{\Upsilon}:\mathbf{P}_{\operatorname{Spin}_{H_{1,3}}^{e}}\left(M\right)\rightarrow M,
\]
and a map
\[
\Upsilon:\mathbf{P}_{\operatorname{Spin}_{H_{1,3}}^{e}}\left(M\right)\rightarrow
\mathbf{P}_{SO_{H_{1,3}}^{e}}\left(M\right),
\]
satisfying the following conditions:
\begin{enumerate}
\item[(a)] $\Pi\left(\Upsilon\left(p\right)\right)=\Pi_{\Upsilon}\left(p\right)$,
$\forall p\in\mathbf{P}_{\operatorname{Spin}_{H_{1,3}}^{e}}\left(M\right)$, where
$\Pi$ is the projection map of the hyperbolic bundle
$\mathbf{P}_{SO_{H_{1,3}}^{e}}\left(M\right)$.

\item[(b)] $\Upsilon\left(pu\right)=\Upsilon\left(p\right)\rho_{u}$,
$\forall p\in\mathbf{P}_{\operatorname{Spin}_{H_{1,3}}^{e}}\left(M\right)$, and
$\rho:\operatorname{Spin}_{H_{1,3}}^{e}\rightarrow\operatorname*{Aut}\left(\mathbb{R}_{1,3}^{H}\right)$,
$\rho_{u}:\mathbb{R}_{1,3}^{H}\rightarrow\mathbb{R}_{1,3}^{H}$ such that
$\rho_{u}x=uxu^{-1}$.
\end{enumerate}
\end{Definition}

\begin{Remark}
The existence of a hyperbolic spinor structure on a spacetime manifold $M$
requires (by analogy with Geroch's classical theorem) the existence of a global
section of the principal bundle $P_{\operatorname{SO}^e_{H_{p,q}}}(M)$. Since
the structure group $\operatorname{SO}^e_{H_{p,q}}$ is strictly larger than
the standard $\operatorname{SO}^e_{1,3}$, one might expect different
topological constraints. However, a detailed analysis of these constraints and
comparison with classical obstruction theory is beyond the scope of this paper
and is left for future investigation.
\end{Remark}

\begin{Remark}
Recall that a local section of the fiber bundle $\left(E,M,\pi,G,F\right)$
on an open set $U\subset M$ is a mapping $S:U\rightarrow E$ such that
$\pi\circ S=I$, and if $U=M$ the section $S$ is said to be global. Then, any
section of the hyperbolic principal fiber bundle
$\mathbf{P}_{\operatorname{Spin}_{H_{1,3}}^{e}}\left(M\right)$ will be called a
hyperbolic spin frame field.
\end{Remark}

\section{Conclusions and Future Directions}\label{sec5}

This paper has presented a comprehensive development of spinor structures
associated with the Hyperbolic Clifford algebra of a real $n$-dimensional
vector space $V$.

\subsection{Summary of Main Results}

The main results are as follows: (i) construction of the frame bundle
$F_H(M) = \cup_{x \in M} F_x M \oplus F_x^* M$ associated to the hyperbolic
tangent-cotangent bundle structure; (ii) derivation of orthogonal
transformations and reflections in hyperbolic space, with explicit
characterization of the Clifford--Lipschitz group $\Gamma_{H_{p,q}}$ as a
surjective cover of $\operatorname{O}_{H_{p,q}}$ with kernel $\mathbb{R}^*$;
(iii) definition and analysis of the Pin groups $\operatorname{Pin}_{H_{p,q}}$
and Spin groups $\operatorname{Spin}_{H_{p,q}}$ as double covers of orthogonal
transformation groups; (iv) establishment of the hyperbolic Clifford bundle
as an associated vector bundle with a reduction in the structure group from
$\operatorname{Aut}(\mathbb{R}_{1,3}^H)$ to $\operatorname{SO}^e_{H_{1,3}}$.

\subsection{Physical Significance}

These results provide a rigorous mathematical foundation for the study of
gravitational fields and particles in theoretical physics, particularly
in formulations that require symmetric treatment of fields and their dual
counterparts. The Clifford bundle formalism offers an elegant alternative
to standard approaches and has proven useful in Yang--Mills type theories
of gravitation in Minkowski spacetime.

We hope this work will inspire further development of the hyperbolic Clifford
algebra formalism and its applications in theoretical physics and algebra.


\begin{thebibliography}{99}

\bibitem{Rodrigues1} Rodrigues, W.A., Jr.; Capelas de Oliveira, E. \textit{The
Many Faces of Maxwell, Dirac and Einstein Equations. A Clifford Bundle
Approach}; Lecture Notes in Physics 722; Springer: New York, NY, USA, 2007.

\bibitem{Sachs} Sachs, R.K.; Wu, H. \textit{General Relativity for
Mathematicians}; Springer: New York, NY, USA, 1977.

\bibitem{Rodrigues3} Rodrigues, W.A., Jr.; Souza, Q.A.G. The Clifford Bundle
and the Nature of the Gravitational Field. \emph{Found. Phys.}
\textbf{1993}, \emph{23}, 1465--1490.

\bibitem{Notte2} Notte-Cuello, E.A.; Rodrigues, W.A., Jr. A Maxwell Like
Formulation of Gravitational Theory in Minkowski Spacetime. \emph{Int.
J. Mod. Phys. D} \textbf{2007}, \emph{16}, 1027--1041.

\bibitem{fr2010} Fern\'{a}ndez, V.V.; Rodrigues, W.A., Jr.
\textit{Gravitation as a Plastic Distortion of the Lorentz Vacuum};
Fundamental Theories of Physics 168; Springer: Berlin, Germany, 2010.

\bibitem{Rodrigues2} Rodrigues, W.A., Jr.; Souza, Q.A.G.; Vaz, J., Jr.
Spinor Fields and Superfields as Equivalence Classes of Exterior
Algebra Fields. In \textit{Clifford Algebras and Spinor Structures};
Ab{\l}amowicz, R., Lounesto, P., Eds.; Kluwer Academic Publishers: Dordrecht,
The Netherlands, 1995; pp. 177--198.

\bibitem{qro7} Rodrigues, W.A., Jr.; Souza, Q.A.G. The Hyperbolic Clifford
Algebra of Multivecfors. \textit{Algebr. Groups Geom.} \textbf{2007},
\emph{24}, 1--24.

\bibitem{Notte1} Notte-Cuello, E.A.; Rodrigues, W.A., Jr. Differential
Structure of the Hyperbolic Clifford Algebra. \emph{Adv. Appl. Clifford
Algebras} \textbf{2015}, \emph{25}, 169--218.

\bibitem{Oziewicz} Oziewicz, Z. From Grassmann to Clifford. In
\textit{Clifford Algebras and their Applications in Mathematical Physics};
Chisholm, J.S.R., Common, A.K., Eds.; D. Reidel Publishing Co.: Dordrecht,
The Netherlands, 1986.

\bibitem{Knus} Knus, M. \emph{Quadratic Forms, Clifford Algebras and Spinors};
Seminars in Mathematics; IMECC--UNICAMP: Campinas, Brazil, 1988.

\bibitem{Lounesto} Lounesto, P. Clifford Algebras and Hestenes Spinors.
\emph{Found. Phys.} \textbf{1993}, \emph{23}, 1203--1237.

\bibitem{Doran} Doran, C.; Lasenby, A. \textit{Geometric Algebra for
Physicists}; Cambridge University Press: Cambridge, UK, 2003.

\bibitem{Todorov} Todorov, I. Clifford Algebras and Spinors. \emph{Bulg. J.
Phys.} \textbf{2011}, \emph{38}, 3--28.

\bibitem{Vaz} Vaz, J., Jr.; da Rocha, R., Jr. \textit{An Introduction to
Clifford Algebras and Spinors}; Oxford University Press: Oxford, UK, 2016.

\bibitem{Osborn} Osborn, H. \textit{Vector Bundles, Volume I}; Academic Press:
New York, NY, USA, 1982.

\bibitem{Geroch1} Geroch, R. Spinor Structure of Space-Times in General
Relativity I. \emph{J. Math. Phys.} \textbf{1968}, \emph{9}, 1739--1744.

\bibitem{Geroch2} Geroch, R. Spinor Structure of Space-Times in General
Relativity II. \emph{J. Math. Phys.} \textbf{1970}, \emph{11}, 343--348.

\bibitem{Lawson} Lawson, H.B., Jr.; Michelsohn, M.-L. \textit{Spin Geometry};
Princeton University Press: Princeton, NJ, USA, 1989.

\bibitem{MonNotte} Notte-Cuello, E.A. \textit{\'{A}lgebras y Fibrados de
Clifford con Aplicaciones}; Editorial Universidad de La Serena: La Serena,
Chile, 2010.

\bibitem{Mosna} Mosna, R.A.; Rodrigues, W.A., Jr. The Bundles of Algebraic
and Dirac--Hestenes Spinor Fields. \emph{J. Math. Phys.} \textbf{2004},
\emph{45}, 2945--2966.

\bibitem{Witten} Witten, E. A Note on the Antibracket Formalism.
\emph{Mod. Phys. Lett. A} \textbf{1990}, \emph{5}, 487--494.

\end{thebibliography}
\end{document}